\documentclass[twocolumn,aps,showpacs,preprintnumbers,amsmath,amssymb,unsortedaddress,superscriptaddress]{revtex4}

\usepackage{graphicx,epsf,epsfig,ulem}

\usepackage{bm}

\usepackage{subfigure}

\usepackage{color}

\usepackage{ulem}

\begin{document}


\title{Phonon runaway in nanotube quantum dots}

\bigskip

\author{L. Siddiqui}
\affiliation{NSF Network for Computational Nanotechnology, Purdue University, West Lafayette, IN 47907}

\author{A. W. Ghosh}
\affiliation{Dept. of Electrical and Computer Engg., University of Virginia, Charlottesville, VA 22903}

\author{S. Datta}
\affiliation{NSF Network for Computational Nanotechnology, Purdue University, West Lafayette, IN 47907}

\medskip

\widetext

\begin{abstract}
We explore electronic transport in a nanotube quantum dot strongly coupled with vibrations and weakly with leads and the thermal environment. We show that the recent observation of anomalous conductance signatures in single-walled carbon nanotube (SWCNT) quantum dots \cite{Dekker,Dekker1} can be understood quantitatively in terms of current driven `hot phonons' that are strongly correlated with electrons. Using rate equations in the many-body configuration space for the joint electron-phonon distribution, we argue that the variations are indicative of strong electron-phonon coupling requiring an analysis beyond the traditional uncorrelated phonon-assisted transport (Tien-Gordon) approach.
\end{abstract}

\maketitle

\section{Introduction}
\noindent One of the significant challenges in microelectronics is controlling the rapidly increasing thermal 
budget associated with current flow through shrinking devices.  
Experimental \cite{Dekker,Dekker1,McEuen,Reed,Natelson,McEuen1,Dai} and theoretical 
\cite{Schoeller,Tuominen,Aji,Flensberg,Millis,Bratkovsky,Nitzan} investigations are revealing intriguing 
aspects of the mutual effect of electronic and vibronic modes on each other. Nanoscale vibrations tend to 
couple strongly with electronic currents and weakly with their `macroenvironment', allowing them to be easily 
driven far from equilibrium. Understanding the dynamics of such electronically driven phonon runaway processes 
is crucial to the evolution of low-power devices, not to mention the novel concepts like molecular motors and 
phonon lasers \cite{rmotor}.

In this paper, we develop a theoretical treatment of current driven nonequilibrium correlated phonon dynamics in nanoscale systems, and use this approach to analyze recent experiments on SWCNT quantum dots (QD) \cite{Dekker}. Using a rate equation for correlated transport in the full many-body eigenspace of the coupled electron-phonon-lead-bath system, we explain novel spectroscopic features such as the anomalously large absorption sidebands arising from phonon runaway in suspended, Coulomb Blockaded nanotubes (Fig. 1).

Our model also explains semi-quantitative features of the experiment such as the amplitude variation of the Coulomb Blockaded conductance peaks and their phonon sidebands, as a function of injected current. However, our model predicts a {\it{linear}} variation in phonon population with current, in contrast 
with experiments that show a {\it{quadratic}} variation \cite{Dekker,Dekker1}. We argue that a possible origin of this discrepancy is because our model explicitly incorporates the effect of strong electron-phonon correlation that is characteristic of these experiments, while the experimentally extracted variation was accomplished by employing a traditional Tien-Gordon analysis that implicitly treats the phonon contribution only through its mean-field oscillating potential acting on the electronic subsystem \cite{Tien}. 

\section{Model}
\noindent We use a model Hamiltonian $H_D$ for a quantum dot with onsite energies $\epsilon_i$, Coulomb interaction energy $U_{ii'}$, vibronic modes at energy $\hbar\omega_j$ and electron-phonon coupling $\lambda_{ij}$ (Fig.~\ref{dot_cont_bath}). The total Hamiltonian including the contacts ($H_C$), the phonon bath ($H_B$) and their couplings with the dot ($H_{DC}$, $H_{DB}$) is: 
\begin{eqnarray}
H &=& H_D+H_C+H_B+H_{DC}+H_{DB}\\
H_D &=& H_{el}+H_{ph}+H_{el-ph}\\
H_{el} &=& \sum_{i}\epsilon_{i}c^{\dagger}_{i}c_i+\frac{1}{2}\sum_{i,i',i\neq i'}U_{ii'}n_{i} n_{i'}\nonumber\\
H_{ph} &=& \sum_{j}\hbar\omega_{j}a^{\dagger}_{j}a_{j}\nonumber\\
H_{el-ph} &=& \sum_{i,j}\lambda_{ij}\hbar\omega_j n_i (a^{\dagger}_{j} +a_{j})\nonumber\\
H_C &=& \sum_{k,\alpha\epsilon{L,R}}\epsilon_{k\alpha}d^{\dagger}_{k\alpha}d_{k\alpha}\\
H_B &=& \sum_{l}\hbar\omega_{l}b^{\dagger}_{l}b_{l}\\
H_{DC} &=& \sum_{i,k,\alpha}\tau_{ik\alpha}(d^{\dagger}_{k\alpha}c_{i}+d_{k\alpha}c^{\dagger}_{i})\\
H_{DB} &=& \sum_{jl}\kappa_{jl}(b^{\dagger}_{l}+b_{l})(a^{\dagger}_{j}+a_{j})
\label{eq.coupling}
\end{eqnarray} where, $c^{\dagger}$ ($c$) and $d^{\dagger}$ ($d$) are the electronic creation (destruction) 
operators for the dot and the leads, $n_{i}=c^{\dagger}_{i}c_{i}$ and $a^{\dagger}$ ($a$) and $b^{\dagger}$ ($b$) 
are the phonon creation (destruction) operators for the dot and the phonon bath respectively. $\tau$, $\epsilon_k$ and $\kappa$ represent respectively the dot-contact coupling, the contact bandstructure and the coupling between the dot vibrations and the thermal bath).

The electron-phonon coupling is eliminated using a standard unitary polaronic transformation $\tilde{H}=
e^{S}He^{-S}$ \cite{Mahan}, where $S=\sum_{i,j}\lambda_{ij}(a^{\dagger}_{j}-a_{j})n_{i}$. This transformation 
renormalizes the onsite ($\tilde{\epsilon}_i$) and Coulomb ($\tilde{U}_{ii'}$) energies for the dot:
\begin{eqnarray}
\tilde{H}_D &=& \sum_{i}\tilde{\epsilon}_{i}\tilde{c}^{\dagger}_{i}\tilde{c}_{i}+\frac{1}{2}\sum_{i,i'}\tilde{U}_{ii'}\tilde{n}_{i}\tilde{n}_{i'}+\sum_{j}\hbar\omega_{j}a^{\dagger}_{j}a_{j}\\
\tilde{\epsilon}_{i} &=& \epsilon_{i}-\sum_{j}\lambda^2_{ij}\hbar\omega_{j}\\
\tilde{U}_{ii'} &=& U_{ii'}-2\sum_{j}\lambda_{ij}\lambda_{i'j}
\label{eq.poltran}
\end{eqnarray} where the `dressed' electron or polaronic annihilation operator
\begin{eqnarray}
\tilde{c}_i &=& e^{S}c_{i}e^{-S} = cX
\label{eq.tranelecoper}
\end{eqnarray} with $X=\textrm{exp}\sum_{j}{\lambda_{ij}(a_{j}-a^{\dagger}_{j})}$ denoting
the generator of the polaronic shift and
\begin{eqnarray}
\tilde{a}_j &=& e^{S}a_{j}e^{-S} = a_{i}-\sum_{j}\lambda_{ij}n_{i}
\label{eq.tranphonoper}
\end{eqnarray} 
while preserving the electronic number operator $\tilde{n}_{i}=n_{i}$. At this point, we will simplify the model 
by considering only a single phonon mode, denoted by the index 1.

\begin{figure}[]
\begin{center}
\includegraphics[width=0.25\textwidth]{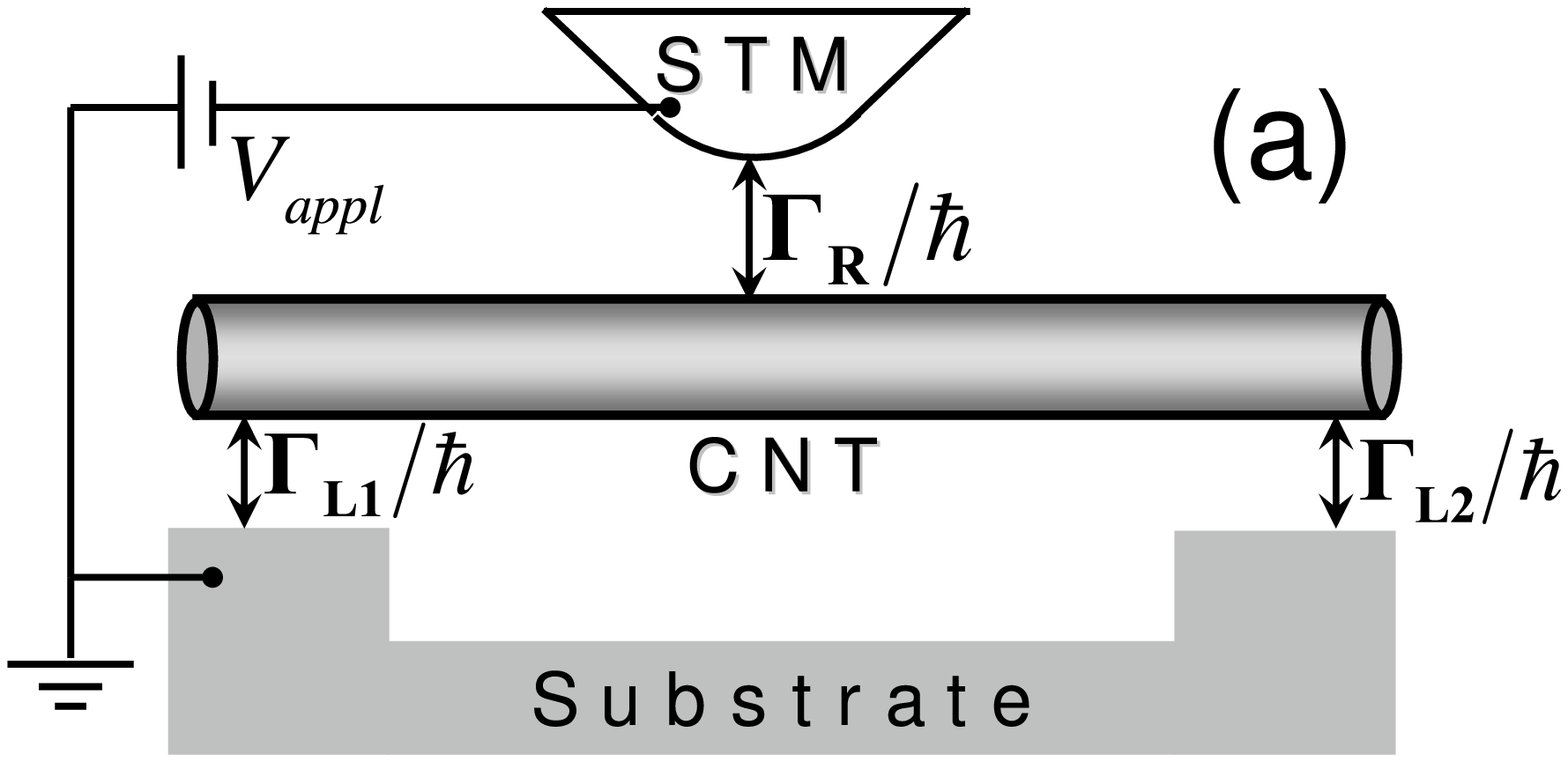}
\end{center}
\begin{center}
\includegraphics[width=0.40\textwidth]{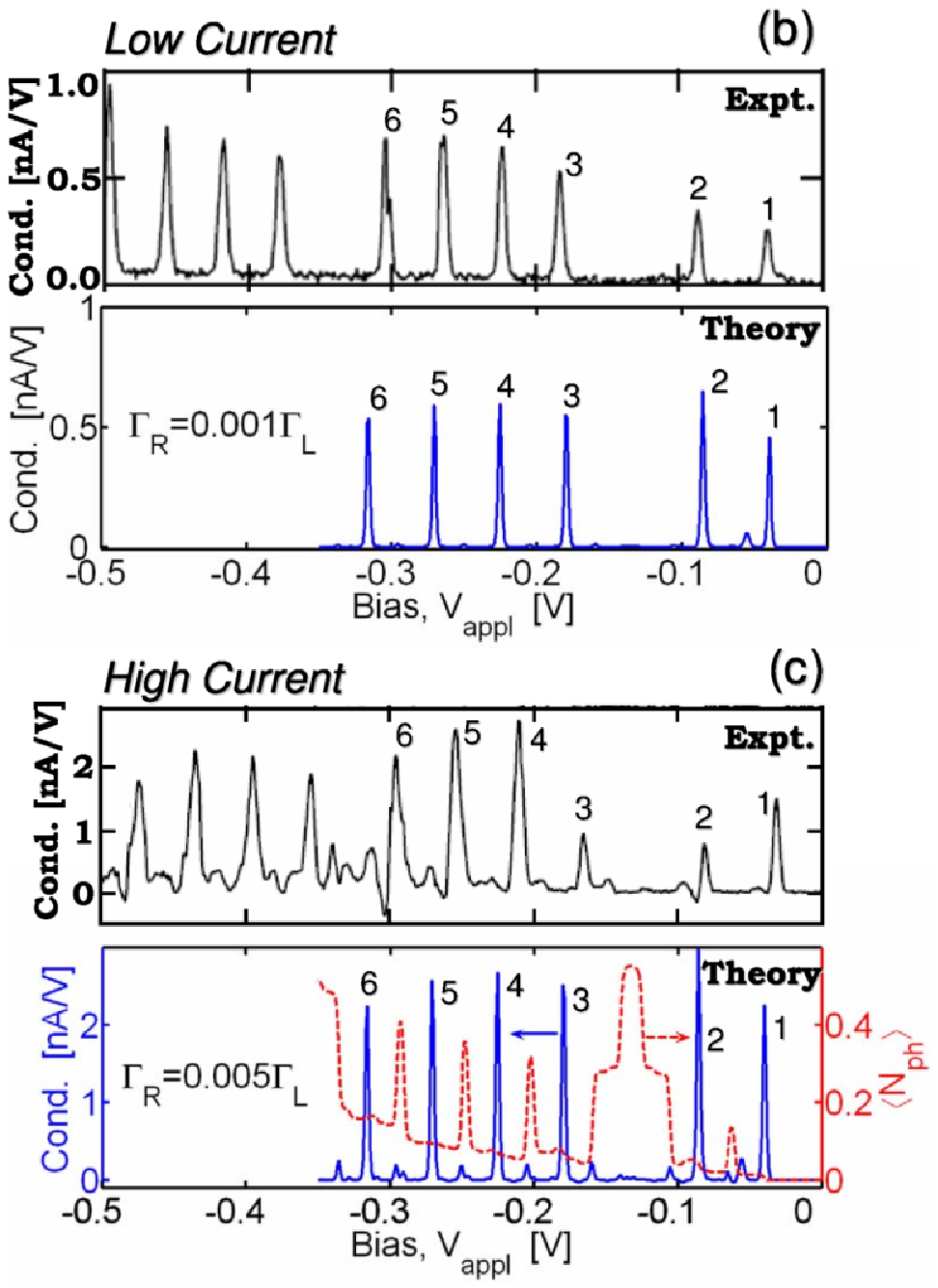}
\end{center}
\caption{(a): Schematic of STM measurement on CNT QD; (b) and (c): Experimental observation and 
theoretical calculation for CNT QD at lower and higher current level ({\it{solid line}}: conductance, 
{\it{dashed line}}: $\langle N_{ph}\rangle$ in (c)). 1,2,..., 6 denotes the main Coulomb peaks. The parameters for the calculation are- $\tilde{\epsilon}_1=\tilde{\epsilon}_2=24$ meV, 
$\tilde{\epsilon}_3=\tilde{\epsilon}_4=\tilde{\epsilon}_5=\tilde{\epsilon}_6=54$ meV, $\tilde{U}=27$ meV, 
$E_F=0$, $\hbar\omega_1=11.5$ meV, $\lambda_{11}=\lambda_{21}=\dots=\lambda_{61}=1.6$, $\Gamma_L=7.5\times 
10^{-6}$ eV, $\beta=5\times 10^{-8}$ eV, $\eta=0.6$ and $T=5$K.}
\label{dekker_data}
\end{figure} 

Current flow in this system involves single electron transitions between many-body states $|e_{Ne}^i,k\rangle$ 
($k$ phonons and {\it{i}}th electronic level in the $N_e$ electronic subspace) of the quantum dot, with matrix 
elements for the electronic destruction operator:
\begin{eqnarray}
&&\langle e_{Ne-1}^r,p|\tilde{c}_i|e_{Ne}^s,k\rangle = \nonumber\\
&&\left\{ \begin{array}{ll}e^{-\lambda^2_{i1}/2}(-\lambda_{i1})^{k-p}\sqrt{\frac{p!}{k!}}\mathcal{L}_p^{k-p}(\lambda^2_{i1}) & \textrm{for $k \ge p$}\\
e^{-\lambda^2_{i1}/2}(-\lambda_{i1})^{p-k}\sqrt{\frac{k!}{p!}}\mathcal{L}_k^{p-k}(\lambda^2_{i1}) & \textrm{for $p \ge k$}
\end{array} \right.
\label{eq.mateldes}
\end{eqnarray} where, $c_{i}|e_{Ne}^s,k>=|e_{Ne-1}^r,k>$ and $\mathcal{L}^{q}_p$ are the associated Laguerre polynomials. For tunneling rates $\Gamma_\alpha/\hbar$ ($\alpha$ = L(left)/R(right)) (Fig.~\ref{dot_cont_bath}), the transition rates are:
\begin{eqnarray}
\mathcal{R}_{|e_{Ne-1}^r,k\rangle\to|e_{Ne}^s,p\rangle}^\alpha= |\langle e_{Ne-1}^r,p|\tilde{c}_i|e_{Ne}^s,k\rangle|^2 \frac{\Gamma_\alpha}{\hbar}\times \nonumber\\
f_\alpha\Biggl({\mathcal{E}} \left(| e^s_{Ne},p\rangle\right)-{\mathcal{E}} \left(|e^r_{Ne-1},k\rangle\right) -\eta |q| V_{appl} \Biggr)
\label{eq.contrate}
\end{eqnarray}

\noindent $f_{L,R}$ are the contact Fermi functions with electrochemical potentials $\mu_L = E_F,~\mu_R = E_F - |q|V_{appl}$ at temperatures $T_{L,R}$, while the electrostatic voltage division factor $\eta$ represents the fraction of the applied bias acting on the levels. Here ${\mathcal{E}}(|\rangle)$ stands for the energy of the eigenstate $|\rangle$ of the dot. 

In our analysis, electrons enter or leave the dot by emitting or absorbing phonons, which in turn are coupled with their thermal environment, held at a bath temperature $T_B$, with an escape rate  $\beta/\hbar$ (Fig.~\ref{dot_cont_bath}), that maintains a Boltzmann ratio between emission and absorption processes in the contact
\begin{eqnarray}
\mathcal{R}^{ph}_{|e^s_{Ne},k\rangle \to |e^s_{Ne},k+1\rangle} &=& \frac{\beta}{\hbar}(k+1)\textrm{exp}[-\frac{\hbar\omega}{k_BT}]\nonumber\\
\mathcal{R}^{ph}_{|e^s_{Ne},k\rangle \to |e^s_{Ne},k-1\rangle} &=& \frac{\beta}{\hbar}k
\label{eq.bathrate}
\end{eqnarray}

The state transitions in the joint electron-phonon many-body space are described by the master equation
\begin{eqnarray}
\frac{dP_{|e^s_{Ne},k\rangle}}{dt}=\sum_{r,N_e',n}\left[P_{|e^r_{Ne'},n\rangle}\mathcal{R}_{|e^r_{Ne'},n\rangle \to |e^s_{Ne},k\rangle}-\right.\nonumber\\
\left.P_{|e^s_{Ne},k\rangle}\mathcal{R}_{|e^s_{Ne},k\rangle \to |e^r_{Ne'},n\rangle}\right]
\label{eq.master}
\end{eqnarray} together with the normalization condition for $P_{|e^s_{Ne},k\rangle}$, where $\mathcal{R}=\mathcal{R}^L+\mathcal{R}^R+\mathcal{R}^{ph}$.

Solving the rate equations at steady state gives us the current $I$ and the steady state population of dot electrons $\langle N_{el}\rangle$ and phonons $\langle N_{ph}\rangle$:
\begin{eqnarray}
I &=& q\sum_{N_e,N_e',r,s,k,n}sgn(N_e'-N_e)P_{|e^s_{Ne},k\rangle}\nonumber\\
&\times &\mathcal{R}^L_{|e^r_{Ne'},n\rangle \to |e^s_{Ne},k\rangle}\nonumber\\
\langle N_{el}\rangle &=& \sum_{s,N_e,k}N_e P_{|e^s_{Ne},k\rangle}\nonumber\\
\langle N_{ph}\rangle &=& \sum_{s,N_e,k}k P_{|e^s_{Ne},k\rangle}
\label{eq.i_nel_nph}
\end{eqnarray} where, $sgn(x)$ is the signum function. This also gives us the phonon generation rate $\mathcal{G}^\alpha$ by the current and the phonon extraction rate $\mathcal{X}^{ph}$ (see appendix) by the bath:
\begin{eqnarray}
\mathcal{G}^{\alpha}&=&\sum_{s,N_e,k}\sum_{r,N_e',n}(n-k)P_{|e^s_{Ne},k\rangle}\mathcal{R}^{\alpha}_{|e^s_{Ne},k> \to |e^r_{Ne'},n\rangle}\nonumber\\
\mathcal{X}^{ph}&=&\sum_{s,N_e,k}\sum_{r,N_e',n}\left(k-n\right)P_{|e^s_{Ne},k\rangle}\mathcal{R}^{ph}_{|e^s_{Ne},k\rangle \to |e^r_{Ne'},n\rangle}\nonumber\\
&=&\left(\frac{\beta}{\hbar}\right)\frac{\langle N_{ph}\rangle-N^{eq}_{ph}}{N^{eq}_{ph}+1}
\label{eq.ph_gen_extr}
\end{eqnarray} where, $N^{eq}_{ph}$ is the equilibrium phonon occupancy at the bath temperature. 

The rate equations are schematically explained in Fig.~\ref{dot_cont_bath}. The inputs to these equations are the electron tunneling rates $\Gamma_{L,R}$, the phonon escape rate $\beta$, the electronic energy configuration $\epsilon_i$, the phonon energy $\hbar\omega_j$, the charging energy $U$ and the electron-phonon coupling $\lambda_{ij}$. In our analysis we will set the lead and bath temperatures $T_{L,R}$ and $T_B$ to be equal to the ambient temperature. 
\begin{figure}[]
\begin{center}
\includegraphics[width=0.45\textwidth]{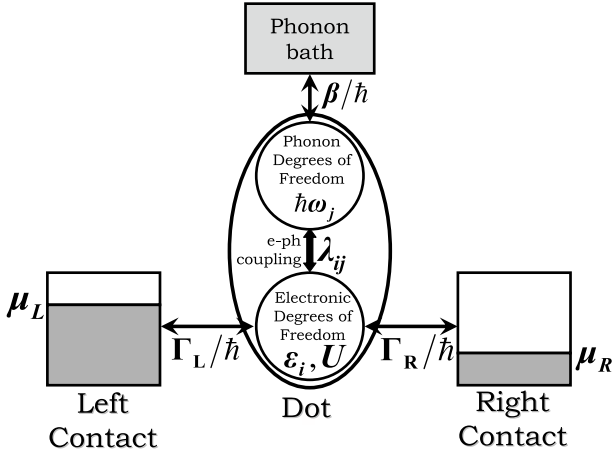}
\end{center}
\caption{The dot is electrically connected to the left (right) contact (with electron tunneling rates $\Gamma_{L,R}/\hbar$) and mechanically to the phonon bath (with a phonon escape rate of $\beta/\hbar$). The dot has electronic degrees of freedom $\epsilon_i$ and phonon degrees of freedom $\hbar\omega_j$ ($i, j=1, 2, 3, \ldots$) with coupling $\lambda_{ij}$.} \label{dot_cont_bath}
\end{figure}

\section{Non-equilibrium phonon occupation in CNT QD}
\subsection{Peaks in the conductance spectrum due to phonon-assisted tunneling}
\noindent We will now apply our many-body rate equations to analyze recent experiments \cite{Dekker,Dekker1} on 
phonon-assisted tunneling in suspended SWCNT quantum dots. The experiment shows several striking features such as 
anomalously large phonon absorption peaks at low temperature (5 K) and a monotonic increase in phonon sideband 
amplitude with current (Fig.~\ref{dekker_data} b and c). Conductance peaks are observed in groups of four, 
suggesting consecutive doubly-spin-degenerate electronic levels in the Coulomb Blockade regime. Sidebands are 
attributed to phonon-assisted tunneling through the radial breathing mode (RBM) of the SWCNT, known to predominate 
at low bias through its effect at the bottom of the conduction band \cite{Avouris}. 

We will model the dot with a single phonon mode ($\hbar\omega_1$) corresponding to the RBM and six electronic energy 
levels, - two lower energy degenerate levels ($\tilde{\epsilon}_1$ and $\tilde{\epsilon}_2$) and four higher energy 
degenerate ones ($\tilde{\epsilon}_3$, $\tilde{\epsilon}_4$, $\tilde{\epsilon}_5$ and $\tilde{\epsilon}_6$). The 
occurrence of the singly-degenerate and the doubly-degenerate discrete electronic levels in CNT QD has been 
observed earlier \cite{Park,Nygard}. We will assume that the tunneling rate $\Gamma_{\alpha}/\hbar$ and the phonon 
escape rate $\beta/\hbar$ are dispersionless, and use the same value for the couplings ($\lambda_{11}=\lambda_{21}=
\dots=\lambda_{61}$) of the RBM to all the electronic levels \cite{rnote}.

In order to match experimental results, we find it necessary to consider electron addition levels 
only, and not electron removal levels lying below the equilibrium Fermi energy 
\cite{Dekker,Dekker1}. An analytical estimate of the height of the first two conductance peaks, 
corresponding to two degenerate levels at $\tilde{\epsilon}_{1,2}$, explains the justification 
behind this assumption. It can be shown that the height of the first two electron {\it{removal}} 
peaks should be proportional to $2\Gamma_{L}\Gamma_{R}/(2\Gamma_{R}+\Gamma_{L})$ and 
$2\Gamma_{L}\Gamma_{R}/(\Gamma_{R}+\Gamma_{L})-2\Gamma_{L}\Gamma_{R}/(2\Gamma_{R}+\Gamma_{L})$ \cite{bmuralid}, assuming that the left contact injects electron\textcolor{blue}, the temperature is very low, 
and ignoring phonon sidebands, - assumptions which are consistent with experimental conditions. However, this 
predicts, for $\Gamma_{R}\ll\Gamma_{L}$, that the height of the second peak should become zero in contrast with the 
experiment (Fig.~\ref{dekker_data}b). The height of the first two electron {\it{addition}} peaks, under the same 
assumption, can be shown to be proportional to $2\Gamma_{L}\Gamma_{R}/(\Gamma_{R}+2\Gamma_{L})$ and $2\Gamma_{L}
\Gamma_{R}/(\Gamma_{R}+\Gamma_{L})-2\Gamma_{L}\Gamma_{R}/(\Gamma_{R}+2\Gamma_{L})$, which predicts, for 
$\Gamma_{R}\ll\Gamma_{L}$, that the peak heights should be equal\textcolor{blue}{,} in agreement with experiment 
(Fig.~\ref{dekker_data}b). This suggests that the observed peaks arise from electron addition rather than 
removal.

Next, let us estimate the input parameters. The phonon energy $\hbar\omega_1$ was measured to be $11.5$ meV 
(see \cite{Dekker,Dekker1}), which together with the separation between the main Coulomb peak and its first 
phonon emission sideband, yields a voltage-division factor $\eta \approx 0.6$. The polaron renormalized charging 
parameter $\tilde{U}$ is estimated to be $30$ meV, as extracted from the separation between consecutive Coulomb peaks originating 
from the same degenerate set of levels. From the estimate of the phonon decay rate and the Q factor, $\beta$ was 
determined to be $\sim 10^{-8}$ eV. In our calculations we varied the tunneling rates $\Gamma_{\alpha}$ and 
electron-phonon coupling constant $\lambda_{i1}$ ($i=1,2,\dots,6$) to match the experimental conductance levels.

Figs.~\ref{dekker_data} b and c show a comparison between the experiment and our calculations with the above 
parameters. Six levels seem sufficient to capture the essential physics, including the increase in number of phonons between emission and absorption sidepeaks arising from the corresponding 
increase or decrease of phonon occupation at those bias points (Fig.~\ref{dekker_data} c). {\it{The calculated phonon 
number significantly exceeds the equilibrium value after each emission peak and drops considerably after absorption, 
indicating strongly correlated nonequilibrium phonon dynamics in this system}}. The weak phonon-substrate coupling 
$\beta$ for suspended tubes leads to a phonon bottleneck whereby the current emits phonons faster than they are 
conducted away, leading to anomalous low temperature absorption peaks that {\it{even exceed}} the corresponding emission peak heights on occasion.

\begin{figure}[]
\begin{center}
\includegraphics[width=0.45\textwidth]{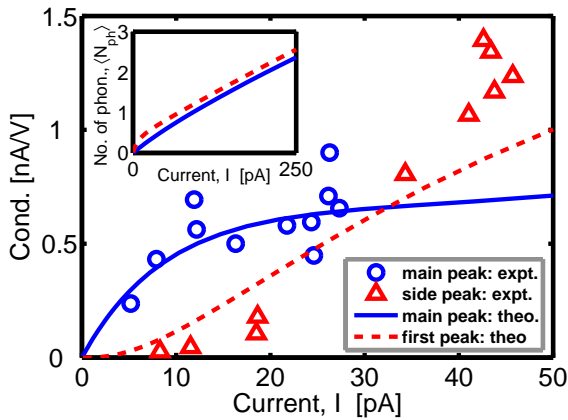}
\end{center}
\caption{Variation of conductance with the current at the second coulomb peak (at $V_{appl}=-0.1$ V) and it's associated first phonon sidepeak (at $V_{appl}=-0.125$ V). {\it{inset}}: Variation of the no. of phonons with current at the above mentioned conductance peaks. The parameters used to generate these results are: $\tilde{\epsilon}_1=\tilde{\epsilon}_2=\tilde{\epsilon}=8$ meV, $\tilde{U}=12$ meV, $E_F=0$, $\hbar\omega_1=12.5$ meV, $\lambda_{11}=\lambda_{21}=2.1$, $\Gamma_L=5\times 10^{-6}$ eV, $\Gamma_R=0.00005\Gamma_L$ to $0.5\Gamma_L$, $\beta=5\times 10^{-10}$ eV, $\eta=0.2$ and $T=5K$.}
\label{Nph_vs_I}
\end{figure}

\subsection{Variation of conductance with current}
\noindent Experiment \cite{Dekker} shows a prominent quadratic dependence of the phonon occupancy on current. This 
variation is extracted indirectly from the observed variation in height of a main conduction peak and it's associated 
first phonon sidepeak with current (Fig.~\ref{Nph_vs_I}), by applying a traditional Tien-Gordon model to fit it 
\cite{Tien}. Note that these reported experimental results are from a different CNT QD sample from the one 
corresponding to Figs.~\ref{dekker_data} b and c. So we will use a model with a different configuration of parameters to capture the essential characteristics of the experimental observations. In order to ascertain the conductance variation with current, we have replotted the reported normalized conductance, $(dI/dV_{appl})/(I/V_{appl})$ vs. current, $I$ to 
conductance $G=dI/dV_{appl}$ vs. current $I$, using the known bias values $V_{appl}$ (-0.1 V at main peak and -0.125 V at first sidepeak) and current values $I$ at those peaks obtained from \cite{Dekker}. The experimental 
observation shows that the main conductance peak initially increases with current and thereafter tends to saturate. 
On the other hand, the conductance at the first side peak gradually increases with current exponentially within the 
entire current range of the experiment. 

\begin{figure}[]
\begin{center}
\includegraphics[width=0.45\textwidth]{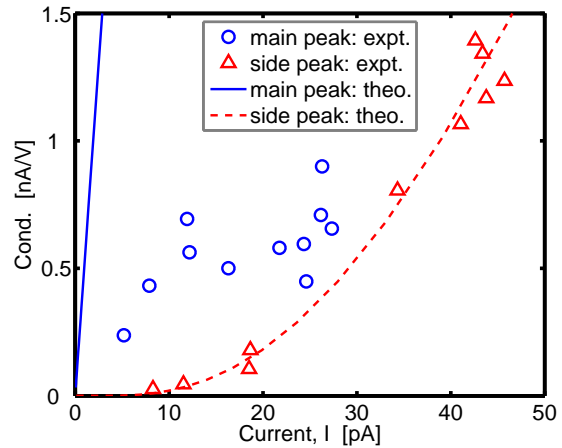}
\end{center}
\caption{Simple one single electron level model do not match the experimental result at all. The values of the 
parameters used are: $\tilde{\epsilon}_1=0.05$ eV, $\hbar\omega_{1}=12.5$ meV, $\lambda_{11}=1.7$, $\Gamma_{L}=
1\times 10^{-5}$ eV, $\Gamma_{R}=0.00005\Gamma_{L}$ to $0.5\Gamma_{L}$, $\beta=5\times 10^{-10}$ eV, $\eta=0.5$ 
and $T=5$ K}
\label{test_one_lev}
\end{figure}

To explore this variation we further simplify our model to just two degenerate single electronic levels at energy 
$\tilde{\epsilon}$. The simulation produces two Coulomb peaks and their associated phonon sidepeaks. To mimic the 
experimental procedure, we increased the current by varying the tunneling rate $\Gamma_{R}$ of the STM tip. The 
calculated current dependences of the conductance at the second main peak (due to the direct tunneling at energy 
$\tilde{\epsilon}+\tilde{U}$) and its associated first phonon sidepeak at higher energy ($\tilde{\epsilon}+\tilde{U}
+\hbar\omega_1$) agree with the experimental trend (Fig.~\ref{Nph_vs_I}). However, we do not find $\langle N_{ph}
\rangle$ varying as $I^{2}$ at the conduction peaks (Fig.~\ref{Nph_vs_I} {\it{inset}}), but instead almost linearly 
within the experimental current range. The disagreement between our result and the claim made in \cite{Dekker} stems from the fundamental difference between our model and the Tien-Gordon model. Unlike our model, the Tien-Gordon model neglects the correlation between the electronic degrees of freedom and the phonon degrees of freedom by replacing the electron-phonon coupling Hamiltonian $H_{\rm{el-ph}}$ with a mean-field oscillating potential $V = \langle \delta H_{\rm{el-ph}}/\delta n\rangle$ which is proportional to the average position operator. For 
a particular phonon mode of frequency $\omega$ and in the absence of leads or baths, the position operator and thus 
the averaged phonon potential is simply proportional to $\cos{\omega t}$. Since the Bohr frequency of the electron 
in turn depends on this oscillating Hamiltonian, the resulting time-evolution operator and thus the electronic 
spectral functions end up with independent Fourier components whose spectral weights depend on Bessel functions, 
in other words, the Tien-Gordon expression. Crucial to the derivation of the Tien-Gordon model is the replacement of 
the electron-phonon coupling Hamiltonian by its mean-field time-dependent component ignoring leads and baths, which 
eliminates all correlation effects. This assumption is clearly inconsistent with our model that shows strongly correlated phonon dynamics, captured by the dimensionless electron-phonon coupling constant $\lambda$ that is typically greater than unity, as mentioned in our figure captions.  

\begin{figure}[]
\begin{center}
\includegraphics[width=0.45\textwidth]{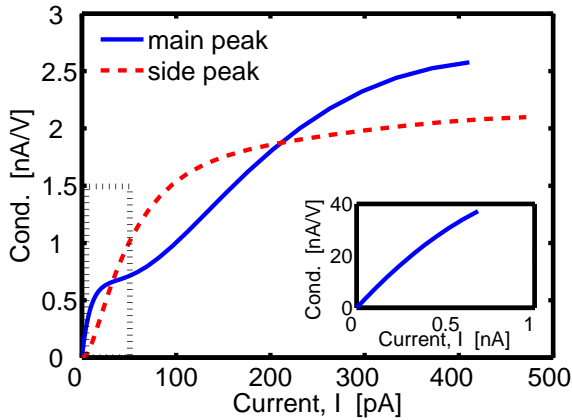}
\end{center}
\caption{Variation of conductance with current for current values higher than the experimental range. {\it{inset}}: The variation of conductance at the coulomb peak in the absence of electron phonon coupling. The values of the parameters used are the same as those of Fig.~\ref{Nph_vs_I} (except that their is no coupling  between electrons and phonons for the results in {\it{inset}}). The region inside the dotted box is the range of experimental observation in Fig.~\ref{Nph_vs_I}.}
\label{full_two_lev}
\end{figure}

It is worth explaining why a single level does not suffice to capture this current variation. We find that the 
conductance of the main Coulomb peak,  irrespective of the parameters chosen, varies linearly with current and 
does not saturate. Since the current prior to the main Coulomb peak is zero, the height of the main conductance peak $\sim I_{coul}/2k_BT$, where $I_{coul}$ is the current plateau immediately past this voltage, the main broadening coming from temperature. This result, consistent with a test simulation with one single electron level (Fig~\ref{test_one_lev}), also persists for the {\it{first}} in a series of Coulomb peaks in a multilevel system for the same reason. In all these examples, the main conductance peak can only vary linearly with the current, in contradiction with experiment, prompting us to look at higher peaks in a more 
complex, multileveled system. This requires us to adopt a more complex model with two single electron levels and to look at the variation at second Coulomb peak and it's associated phonon sidepeak. The results reported in the 
experiment (Figs. 4a and 4b of \cite{Dekker}) were later confirmed to involve the variation of second Coulomb peak and its associated phonon sidepeak \cite{DDekker}.

We used our doubly degenerate single electron level model to explore the variation of the conductance for current 
values beyond the experimental current range, to project the characteristics
under higher STM set current. The results are shown in Fig.~\ref{full_two_lev}. The conductance values at both the Coulomb peak and the first sidepeak show a non-trivial variation with current. The variation of the conductance at the Coulomb peak is compared to that in (Fig.~\ref{full_two_lev} {\it{inset}}) absence of electron-phonon coupling. We find that the coupling of electronic 
and phonon degrees of freedom significantly affects the Coulomb peak characteristics. In particular, while the 
trends match the experimental data up to the experimental current levels, we notice that at higher currents the main 
peak goes through a kink followed by a rise\textcolor{blue}{,} while the sidepeak amplitude tends to saturate.

\begin{figure}[]
\begin{center}
\includegraphics[width=0.45\textwidth]{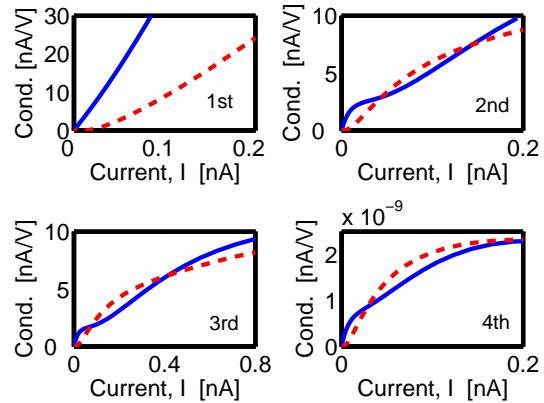}
\end{center}
\caption{Variation of conductance with current for four degenerate single electron levels at the four Coulomb peaks and their associated first phonon sidepeaks. The values of the parameters used are: $\tilde{\epsilon}_1=\tilde{\epsilon}_2=\tilde{\epsilon}_3=\tilde{\epsilon}_4=\tilde{\epsilon}=20$ meV, $\tilde{U}=30$ meV, $E_F=0$, $\hbar\omega_1=12.5$ meV, $\lambda_{11}=\lambda_{21}=\lambda_{31}=\lambda_{41}=2.1$, $\Gamma_L=5\times 10^{-6}$ eV, $\Gamma_R=0.00005\Gamma_L$ to $0.5\Gamma_L$, $\beta=5\times 10^{-10}$ eV, $\eta=0.5$ and $T=5K$.}
\label{full_four_lev}
\end{figure}

Finally we look at the conductance variation with current for a quadruply degenerate single electron level model for a CNT QD that occasionally shows doubly spin degenerate shells. The conductance variation at each Coulomb peak and its associated first higher energy phonon sidepeak are shown in Fig.~\ref{full_four_lev}. The first Coulomb peak varies almost linearly with current (similar to 
Fig.~\ref{test_one_lev}), as we argued earlier. The conductance variations at the other three peaks show the same qualitative trends as Fig.~\ref{full_two_lev}. From the qualitative similarity between Figures \ref{test_one_lev}, \ref{full_two_lev} and \ref{full_four_lev}, we see that the qualitative features of these variations are robust with respect to the values of the parameters that went into the model.

\section{Conclusion}
\noindent We have studied strong electron-phonon coupled dynamics under nonequilibrium conditions using a 
many-electron rate equation approach, focusing on recent experiments on suspended nanotubes. From our calculations 
we argue that phonons in weakly coupled systems can be readily driven far from equilibrium, leading to anomalous 
signatures in the corresponding conductance spectrum and temperature dependences. Interesting extensions of this 
work would involve studying the imputations of nonequilibrium phonon dynamics for energy dissipation in nanoscale 
systems, and of maintaining phonon coherence by tuning their decay rates through their mechanical coupling with the 
substrate, and through phonon-phonon interactions controlled by the inherent nonlinearity of the lattice.

\section{Acknowledgment}
\noindent We would like to thank Prof. Cees Dekker and Dr. Brian LeRoy for sharing electronic data and for their valuable comments on an earlier version of this manuscript. We would also like to acknowledge NSF Network for Computational Technology (NCN) for providing the computational resources. This work was supported by DURINT and DARPA.

\section{Appendix}
In this section we show the derivation of some important analytical results. The derivations in \ref{sec:tranelecoper} and \ref{sec:diagham}, which are already dicussed in ref.~\cite{Bratkovsky,Mahan}, have been worked out here to maintain continuity.

\subsection{Derivation of Equation~\ref{eq.tranelecoper}\label{sec:tranelecoper} and~\ref{eq.tranphonoper}}
\noindent The terms like $e^{S}Ae^{-S}$ are evaluated using:
\begin{eqnarray}
e^{S}Ae^{-S}=A+[S,A]+(1/2!)[S,[S,A]]+\dots
\label{eq.P}
\end{eqnarray}
\noindent When $A=c_{i}$ we have:
\begin{eqnarray}
[S,c_{i}] &=& \sum_{h,j}\lambda_{hj}(a^{\dagger}_{j}-a_{j})[c^{\dagger}_{h}c_{h},c_{i}]\nonumber\\
&=& \sum_{h,j}\lambda_{hj}(a^{\dagger}_{j}-a_{j})[c^{\dagger}_{h}c_{h}c_{i}-c_{i}c^{\dagger}_{h}c_{h}]\nonumber\\
&=& \sum_{h,j}\lambda_{hj}(a^{\dagger}_{j}-a_{j})[-c^{\dagger}_{h}c_{i}c_{h}-c_{i}c^{\dagger}_{h}c_{h}]\nonumber\\
&=& \sum_{h,j}\lambda_{hj}(a^{\dagger}_{j}-a_{j})[(c_{i}c^{\dagger}_{h}-\delta_{i,h})c_{h}-c_{i}c^{\dagger}_{h}c_{h}]\nonumber\\
&=& \sum_{h,j}\lambda_{hj}(a^{\dagger}_{j}-a_{j})[-\delta_{i,h}c_{h}]\nonumber\\
\left[S,c_{i}\right] &=& \sum_{i,j}\lambda_{ij}(a_{j}-a^{\dagger}_{j})c_{i}
\label{eq.Q}
\end{eqnarray} where $\delta$ is the kronecker delta function. Then,
\begin{eqnarray}
[S,[S,c_{i}]] &=& [S,\sum_{i,j}\lambda_{ij}(a^{\dagger}_{j}-a_{j})c_{i}]\nonumber\\
&=& \sum_{i,j}\left[\lambda_{ij}(a^{\dagger}_{j}-a_{j})\right]\sum_{h,j}\left[\lambda_{hj}(a^{\dagger}_{j}-a_{j})\right][c^{\dagger}_{h}c_{h},c_{i}]\nonumber\\
\left[S,[S,c_{i}]\right] &=& \left[\sum_{i,j}\lambda_{ij}(a_{j}-a^{\dagger}_{j})\right]^{2}c_{i}
\label{eq.R}
\end{eqnarray} and so on. Putting all these results in equation ~\ref{eq.P} we finally get eq.~\ref{eq.tranelecoper}.
\noindent When $A=a_{j}$ we have:
\begin{eqnarray}
[S,a_{j}] &=& \sum_{i,h}n_{i}\lambda_{ih}[a^{\dagger}_{h}-a_{h},a_j]\nonumber\\
&=& \sum_{i,h}n_{i}\lambda_{ih}(a_{h}a_{j}-a^{\dagger}_{h}a_{j}-a_{j}a_{h}+a_{j}a^{\dagger}_{h})\nonumber\\
&=& \sum_{i,h}n_{i}\lambda_{ih}\delta_{j,h}\nonumber\\
\left[S,a_{j}\right] &=& \sum_{i}n_{i}\lambda_{ij}
\label{eq.S}
\end{eqnarray} So, afterwards $[S,[S,a_{j}]]=[S,[S,[S,a_{j}]]]=\dots=0$. Putting all these commutators in equation~\ref{eq.P} we get eq.~\ref{eq.tranphonoper}

\subsection{Derivation of Equation 7, 8 and 9}\label{sec:diagham}
\noindent Applying the polaronic transformation:
\begin{eqnarray}
\tilde{H}_{D} &=& \tilde{H}_{el}+\tilde{H}_{ph}+\tilde{H}_{el-ph}
\label{eq.T}
\end{eqnarray} Now the electronic portion:
\begin{eqnarray}
\tilde{H}_{el} &=& e^{S}H_{el}e^{-S}\nonumber\\
\tilde{H}_{el} &=& \sum_{i}\epsilon_{i}\tilde{n}_{i}+\frac{1}{2}\sum_{i,i',i\neq i'}U_{ii'}\tilde{n}_{i}\tilde{n}_{i'}
\label{eq.U}
\end{eqnarray} The phonon portion:
\begin{eqnarray}
\tilde{H}_{ph} &=& e^{S}H_{ph}e^{-S}\nonumber\\
\tilde{H}_{ph} &=& \sum_{j}\hbar\omega_j\tilde{a}^{\dagger}_{j}a_{j}\nonumber\\
&=& \sum_{j}\hbar\omega_{j}(a^{\dagger}_{j}-\sum_{i}\lambda_{ij}n_{i})(a^{\dagger}_{j}-\sum_{i'}\lambda_{i'j}n_{i'})\nonumber\\
&=& \sum_{j}\hbar\omega_{j}a^{\dagger}_{j}a_{j}+\sum_{i,i',j}\hbar\omega_{j}\lambda_{ij}\lambda_{i'j}n_{i}n_{i'}\nonumber\\
&-&\sum_{i,j}\hbar\omega_{j}\lambda_{ij}n_{i}a_{j}-\sum_{i',j}\hbar\omega_{j}\lambda_{i'j}n_{i'}a^{\dagger}_{j}\nonumber\\
\tilde{H}_{ph} &=& \sum_{j}\hbar\omega_{j}a^{\dagger}_{j}a_{j}+\sum_{i,i',j,i\neq i'}\hbar\omega_{j}\lambda_{ij}\lambda_{i'j}\tilde{n}_{i}\tilde{n}_{i'}\nonumber\\
&+&\sum_{i,j}\lambda^{2}_{ij}\hbar\omega_{j}\tilde{n}_{i}-\sum_{i,j}\hbar\omega_{j}\lambda_{ij}\tilde{n}_{i}(a^{\dagger}_{j}+a_{j})
\label{eq.V}
\end{eqnarray} where we have used $n_{i}n_{i}=n_{i}=\tilde{n}_{i}$. Now, finally, the electron phonon coupling term:
\begin{eqnarray}
\tilde{H}_{el-ph} &=& e^{S}H_{el-ph}e^{-S}\nonumber\\
\tilde{H}_{el-ph} &=& \sum_{i,j}\lambda_{ij}\hbar\omega_{j}\tilde{n}_{i}(\tilde{a}^{\dagger}_{j}+\tilde{a}_{j})\nonumber\\
&=& \sum_{i,j}\hbar\omega_{j}\lambda_{ij}\tilde{n}_{i}(a^{\dagger}_{j}+a_{j}-\sum_{i'}\lambda_{i'j}n^{\dagger}_{i'}-\sum_{i'}\lambda_{i'j}n_{i'})\nonumber\\
&=& \sum_{i,j}\hbar\omega_{j}\lambda_{ij}\tilde{n}_{i}(a^{\dagger}_{j}+a_{j})-2\sum_{i,i',j}\hbar\omega_{j}\lambda_{ij}\lambda_{i'j}\tilde{n}_{i}\tilde{n}_{i'}\nonumber\\
\tilde{H}_{el-ph} &=& \sum_{i,j}\hbar\omega_{j}\lambda_{ij}\tilde{n}_{i}(a^{\dagger}_{j}-2\sum_{i,j}\lambda^{2}_{ij}\hbar\omega_{j}\tilde{n}_{i}\nonumber\\
&-&2\sum_{i,i',j,i\neq i'}\hbar\omega_{j}\lambda_{ij}\lambda_{i'j}\tilde{n}_{i}\tilde{n}_{i'}
\label{eq.W}
\end{eqnarray} Substituting equations~\ref{eq.U},~\ref{eq.V} and~\ref{eq.W} in equation~\ref{eq.T} we get the results of equations 7, 8 and 9.

\subsection{Derivation of Equation~\ref{eq.mateldes}}
We will start from:
\begin{eqnarray}
\langle e_{Ne-1}^r,p|\tilde{c}_i|e_{Ne}^s,k\rangle &=& \langle e_{Ne-1}^r,p|c_{i}X|e_{Ne}^s,k\rangle\nonumber
\end{eqnarray} Now since $c_{i}|e_{Ne}^s,p>=|e_{Ne-1}^r,p>$ it follows that:
\begin{eqnarray}
\langle e_{Ne-1}^r,p|\tilde{c}_i|e_{Ne}^s,k\rangle &=& \langle e_{Ne}^s,p|X|e_{Ne}^s,k\rangle\nonumber
\end{eqnarray} For one phonon mode $X=\textrm{exp}[\lambda_{i1}(a_{1}-a^{\dagger}_{1})]$. Then:
\begin{eqnarray}
\langle e_{Ne-1}^r,p|\tilde{c}_i|e_{Ne}^s,k\rangle &=& \langle e_{Ne}^s,p|e^{\lambda_{i1}(a_{1}-a^{\dagger}_{1})}|e_{Ne}^s,k\rangle\nonumber
\end{eqnarray} Using $e^{\lambda_{i1}(a_{1}-a^{\dagger}_{1})}=e^{-\lambda_{i1}a^{\dagger}_{1}}e^{\lambda_{i1}a_{1}}e^{-\lambda^{2}_{i1}/2}$:
\begin{eqnarray}
&\langle &e_{Ne-1}^r,p|\tilde{c}_i|e_{Ne}^s,k\rangle =\nonumber\\
&&e^{-\lambda^{2}_{i1}/2}\langle e_{Ne}^s,p|e^{-\lambda_{i1}a^{\dagger}_{1}}e^{\lambda_{i1}a_{1}}|e_{Ne}^s,k\rangle\nonumber\\
\label{eq.A}
\end{eqnarray} By simple algebra:
\begin{eqnarray}
&&e^{\lambda_{i1}a_{1}}|e_{Ne}^s,k\rangle = \sum^{\infty}_{l=0}\frac{(\lambda_{i1})^{l}}{l!}(a_{1})^{l}|e_{Ne}^s,k\rangle\nonumber\\
&&e^{\lambda_{i1}a_{1}}|e_{Ne}^s,k\rangle = \nonumber\\
&&\sum^{k}_{l=0}\frac{(\lambda_{i1})^{l}}{l!}\left[\frac{k!}{(k-l)!}\right]^{1/2}|e_{Ne}^s,k-l\rangle
\label{eq.B}
\end{eqnarray} and similarly
\begin{eqnarray}
&&\langle e_{Ne}^s,p|e^{-\lambda_{i1}a^{\dagger}_{1}} = \nonumber\\
&&\sum^{p}_{m=0}\langle e_{Ne}^s,p-m|\frac{(-\lambda_{i1})^{m}}{m!}\left[\frac{p!}{(p-m)!}\right]^{1/2}
\label{eq.C}
\end{eqnarray} From equations ~\ref{eq.A},~\ref{eq.B} and ~\ref{eq.C},
\begin{eqnarray}
\langle e_{Ne-1}^r,p|&\tilde{c}_i&|e_{Ne}^s,k\rangle = e^{-\lambda^{2}_{i1}/2}\sum^{p}_{m=0}\sum^{k}_{l=0}\left[\frac{(\lambda_{i1})^{l}(-\lambda_{i1})^{m}}{m!l!}\right.\nonumber\\
&\times &\left.\frac{(p!k!)^{1/2}}{{(p-m)!(k-l)!}^{1/2}}\delta_{p-m,k-l}\right]
\label{eq.D}
\end{eqnarray}
\noindent For the case of $k\ge p$, to remove the kronecker delta function, we substitute $l=k-p+m$ in Eq. ~\ref{eq.D}:
\begin{eqnarray}
\langle e_{Ne-1}^r,p|&\tilde{c}_i&|e_{Ne}^s,k\rangle = e^{-\lambda^{2}_{i1}/2}\sum^{p}_{m=0}\left[\frac{(\lambda_{i1})^{k-p+m}(-\lambda_{i1})^{m}}{m!(k-p+m)!}\right.\nonumber\\
&\times &\left.\frac{(p!k!)^{1/2}}{(p-m)!}\right]\nonumber\\
&=&e^{-\lambda^{2}_{i1}/2}\sum^{p}_{m=0}\frac{(-1)^{m}(\lambda_{i1})^{k-p}(\lambda_{i1}^{2})^{m}(k!p!)^{1/2}}{m!(k-p+m)!(p-m)!}\nonumber\\
&=&e^{-\lambda^{2}_{i1}/2}(\lambda_{i1})^{k-p}\sqrt{\frac{p!}{k!}}\nonumber\\
&\times &\sum^{p}_{m=0}\frac{(-1)^{m}\{(k-p)+p\}!(\lambda_{i1}^{2})^{m}}{m!\{(k-p)+m\}!(p-m)!}\nonumber
\end{eqnarray} The summed series is nothing but associated Laguerre polynomial $\mathcal{L}^{k-p}_{p}(\lambda^{2}_{i1})$. So, finally we get:
\begin{eqnarray}
&\langle &e_{Ne-1}^r,p|\tilde{c}_i|e_{Ne}^s,k\rangle = \nonumber\\
&&\begin{array}{ll}
e^{-\lambda^2_{i1}/2}(\lambda_{i1})^{k-p}\sqrt{\frac{p!}{k!}}\mathcal{L}_p^{k-p}(\lambda^2_{i1}) & \textrm{for $k \ge p$}\nonumber
\end{array}
\end{eqnarray} Following the exact same procedure we can show that,
\begin{eqnarray}
&\langle &e_{Ne-1}^r,p|\tilde{c}_i|e_{Ne}^s,k\rangle = \nonumber\\
&&\begin{array}{ll} e^{-\lambda^2_{i1}/2}(\lambda_{i1})^{p-k}\sqrt{\frac{k!}{p!}}\mathcal{L}_k^{p-k}(\lambda^2_{i1}) & \textrm{for $p \ge k$}\nonumber
\end{array}
\end{eqnarray}

\subsection{Derivation of Equation~\ref{eq.ph_gen_extr}}
From the equations ~\ref{eq.bathrate}, we can see that the phonon bath induces a transition between only those two states which have the exact same electronic configuration and phonon number differing by 1. So we can write:
\begin{eqnarray}
\mathcal{X}^{ph}&=&\sum_{s,Ne,k}\sum_{r,Ne',n}\left(k-n\right)P_{|e^s_{Ne},k\rangle}\mathcal{R}^{ph}_{|e^s_{Ne},k\rangle \to |e^r_{Ne'},n\rangle}\nonumber\\
&=&\sum_{s,N_e,k}\left(k-k\mp 1\right)P_{|e^s_{Ne},k\rangle}\mathcal{R}^{ph}_{|e^s_{Ne},k\rangle \to |e^s_{Ne},k\pm 1\rangle}\nonumber\\
&=&\sum_{s,N_e}\sum_{k=1}^{\infty}P_{|e^s_{Ne},k\rangle}\mathcal{R}^{ph}_{|e^s_{Ne},k\rangle \to |e^s_{Ne},k-1\rangle}\nonumber\\
&-&\sum_{s,N_e}\sum_{k=0}^{\infty}P_{|e^s_{Ne},k\rangle}\mathcal{R}^{ph}_{|e^s_{Ne},k\rangle \to |e^s_{Ne},k+1\rangle}\nonumber\\
&=&\frac{\beta}{\hbar}\sum_{s,N_e}\sum_{k=0}^{\infty}P_{|e^s_{Ne},k\rangle}\left[k-(k+1)\textrm{exp}[-\frac{\hbar\omega}{k_BT}]\right]\nonumber\\
&=&\left(\frac{\beta}{\hbar}\right)\left(1-\textrm{exp}[-\frac{\hbar\omega}{k_BT}]\right)\sum_{s,N_e}\sum_{k=0}^{\infty}kP_{|e^s_{Ne},k\rangle}\nonumber\\
&-&\left(\frac{\beta}{\hbar}\right)\textrm{exp}[-\frac{\hbar\omega}{k_BT}]\sum_{s,N_e}\sum_{k=0}^{\infty}P_{|e^s_{Ne},k\rangle}\nonumber
\end{eqnarray} So, from Eq. ~\ref{eq.i_nel_nph} and the normalization condition:
\begin{eqnarray}
\mathcal{X}^{ph}&=&\left(\frac{\beta}{\hbar}\right)\left[\langle N_{ph}\rangle\left(1-\textrm{exp}[-\frac{\hbar\omega}{k_BT}]\right)-\textrm{exp}[-\frac{\hbar\omega}{k_BT}]\right]\nonumber\\
&=&\left(\frac{\beta}{\hbar}\right)\left[\langle N_{ph}\rangle\left(1-\frac{N^{eq}_{ph}}{N^{eq}_{ph}+1}\right)-\frac{N^{eq}_{ph}}{N^{eq}_{ph}+1}\right]\nonumber\\
\mathcal{X}^{ph}&=&\left(\frac{\beta}{\hbar}\right)\frac{\langle N_{ph}\rangle-N^{eq}_{ph}}{N^{eq}_{ph}+1}\nonumber
\end{eqnarray} where, $N^{eq}_{ph}=\left[\textrm{exp}\right[\hbar\omega/(k_{B}T)]-1]^{-1}$.


\begin{thebibliography}{100}
\bibitem{rnote} Note that assumptions of uniform electron-phonon coupling and charging energy can be easily relaxed, as demonstrated in \cite{Wegewijs}.
\end{thebibliography}
\end{document}